\documentclass[%
reprint,
superscriptaddress,
amsmath,amssymb,
aps,
floatfix,
]{revtex4-1}
\usepackage{graphicx}% Include figure files
\usepackage{dcolumn}% Align table columns on decimal point
\usepackage{bm}% bold math
\usepackage{hyperref}% add hypertext capabilities
\usepackage[usenames]{color}
\usepackage[dvipsnames]{xcolor}
\begin{document}

\preprint{APS/123-QED}

\title{Inverse Compton emission from millisecond pulsars in the Galactic bulge}% Force line breaks with \\

\author{Deheng Song}\email{dhsong@vt.edu}
\affiliation{Center for Neutrino Physics, Department of Physics, Virginia Tech, Blacksburg, Virginia 24061, USA}
\author{Oscar Macias}\email{oscar.macias@ipmu.jp}
\affiliation{Center for Neutrino Physics, Department of Physics, Virginia Tech, Blacksburg, Virginia 24061, USA}
\affiliation{Kavli Institute for the Physics and Mathematics of the Universe (WPI), University of Tokyo, Kashiwa, Chiba 277-8583, Japan}
\affiliation{GRAPPA Institute, University of Amsterdam, 1098 XH Amsterdam, Netherlands}
\author{Shunsaku Horiuchi}\email{horiuchi@vt.edu}
\affiliation{Center for Neutrino Physics, Department of Physics, Virginia Tech, Blacksburg, Virginia 24061, USA}

\date{Received 21 January 2019; published 25 June 2019}% It is always \today, today,
% but any date may be explicitly specified
\begin{abstract}
Analyses of Fermi Gamma-Ray Space Telescope data have revealed a source of excess diffuse gamma rays towards the Galactic center that extends up to roughly $\pm$20 degrees in latitude. The leading theory postulates that this GeV excess is the aggregate emission from a large number of faint millisecond pulsars (MSPs). The electrons and positrons ($e^\pm$) injected by this population could produce detectable inverse-Compton (IC) emissions by up-scattering ambient photons to gamma-ray energies. In this work, we calculate such IC emissions using \texttt{GALPROP}. A triaxial three-dimensional model of the bulge stars obtained from a fit to infrared data is used as a tracer of the putative MSP population. This model is compared against one in which the MSPs are spatially distributed as a Navarro-Frenk-White squared profile. We show that the resulting spectra for both models are indistinguishable, but that their spatial morphologies have salient recognizable features. The IC component above $\sim$TeV energies carries information on the spatial morphology of the injected $e^\pm$. Such differences could potentially be used by future high-energy gamma-ray detectors such as the Cherenkov Telescope Array to provide a viable multiwavelength handle for the MSP origin of the GeV excess.
\end{abstract}
\maketitle
\section{Introduction}
In the past decade, the Fermi Large Area Telescope (Fermi-LAT) has provided accurate observations of the gamma-ray sky, of which the Galactic center (GC) remains one of the most intriguing and intricate regions. It is of paramount importance to map the gamma-ray emissions from the GC to better understand the properties of cosmic rays (CRs), the interstellar medium (ISM), and tests of dark matter (DM) in the inner regions of our Galaxy. Using template fitting techniques to regress out the Galactic and extragalactic diffuse emissions, multiple studies~\cite{Goodenough:2009gk,Vitale:2009hr,Hooper:2010mq,Abazajian:2012pn,Gordon:2013vta, Macias:2013vya,Hooper:2013rwa,Abazajian:2014fta,Daylan:2014rsa,Calore:2014xka,Zhou:2014lva,TheFermi-LAT:2015kwa,TheFermi-LAT:2017vmf} have found an excess of gamma rays towards the GC extending up to $\sim\pm$20 degrees in latitude. Often referred to as the Galactic center excess (GCE), this excess emission has a centrally peaked spatial morphology that is roughly spherically symmetric with a radial power law of slope $\sim$2.4, and a curved energy spectrum peaking at $\sim$3 GeV. Some authors have argued that the GCE is consistent with a DM emission~\cite{Goodenough:2009gk,Abazajian:2012pn,Gordon:2013vta,Macias:2013vya,Calore:2014xka,Daylan:2014rsa} given its similarities to simplified predictions of weakly interacting massive particle DM models.

However, studies have also shown that the origin of the GCE can be explained by astrophysical sources, such as a population of unresolved gamma-ray-emitting millisecond pulsars (MPSs)~\cite{Abazajian:2010zy,Abazajian:2012pn,Gordon:2013vta,Macias:2013vya,Calore:2014xka,Daylan:2014rsa}. While there is ongoing debate regarding the consistency of the GCE with the luminosity function of MSPs measured elsewhere in the Galaxy\cite{Cholis:2014lta, Hooper:2015jlu, Ploeg:2017vai, Bartels:2018xom}, evidence supporting the MSP hypothesis is mounting. For example, population synthesis simulations~\cite{Gonthier:2018ymi} show that $\sim$$10^4$ MSPs can inhabit the GC region and explain the GCE. Also, subthreshold photon-count statistics have shown detectable features that can be used to distinguish between the MSPs and DM interpretations of the GCE. In particular, Refs.~\cite{Lee:2015fea,Mishra-Sharma:2016gis} introduced a new statistical technique called the non-Poissonian template fit which can be used for characterizing populations of unresolved point sources at fluxes below the detection threshold. They have shown that an unresolved population of point sources just below the sensitivity of the LAT is responsible for the GCE.

More recently, Refs.~\cite{Macias:2016nev, Bartels:2017vsx,  Macias:2019omb} investigated whether the spatial morphology of the GCE is better described by the distribution of stars or by DM. It has been firmly established that the bulk of the stars in the GC region form a so-called box/peanut-bulge structure~\cite{Dwek:1995xu,Freudenreich:1997bx,LopezCorredoira:1999dg,Nataf:2010wf,Wegg:2013upa}. The density distribution of these bulge stars can be reasonably described by a triaxial geometric function~\cite{Freudenreich:1997bx,Nataf:2010wf,Wegg:2013upa} extending in length to a few kpc from the GC. In addition to the box/peanut-bulge, there is a distinct stellar population in the innermost $\sim$200 pc of the Galaxy called the nuclear bulge (NB)~\cite{Nishiyama:2013eba}. Reference~\cite{Macias:2016nev} used two different stellar maps for the box/peanut-bulge~\cite{Freudenreich:1997bx,Ness:2016aaa} and also considered the NB of Ref.~\cite{Nishiyama:2013eba}, and demonstrated that such nonspherical bulge morphologies provide a significantly better fit to the GCE than a spherical Navarro-Frenk-White squared (NFW$^2$) spatial map describing a DM annihilation signal. Importantly, that article showed that once a bulge model is included in the analysis, there is no longer statistically significant evidence for a NFW$^2$ component. These results have been corroborated by Ref.\cite{Bartels:2017vsx} using a different and more flexible analysis method~\cite{Storm:2017arh} and by Ref.~\cite{Macias:2019omb} using an improved Galactic diffuse emission model.

The population of MSPs in the Galactic bulge would not only produce prompt gamma-ray emission correlated with their spatial distribution, but also inject $e^\pm$ into the interstellar environment. These CR $e^\pm$ can produce secondary emissions by interacting with the ISM and magnetic field. It has been pointed out that while the prompt gamma rays from MSPs are expected to follow the morphology of the source distribution, secondary emissions---inverse Compton (IC), bremsstrahlung, and synchrotron radiation---are expected to have different morphologies, since they also depend on their relevant targets, i.e., the interstellar radiation field (ISRF), the gas distribution, and the magnetic field of the Galaxy, respectively. The IC component is a result of an energy-dependent convolution of the spatial morphology of the CR $e^\pm$ sources with the ambient photon fields from starlight, infrared (IR) light, and the cosmic microwave background (CMB).

Previous works~\cite{Yuan:2014yda,Petrovic:2014xra} have studied the secondary IC emission at $\sim$GeV-TeV energies from MSP $e^\pm$ interacting with the ISRF assuming a \emph{spherically symmetric} distribution of MSPs. Under such an assumption, Ref.~\cite{Lacroix:2015wfx} searched for secondary gamma-ray emission from MSPs by performing template fits to the GCE data, and found it to be difficult to detect or constrain the putative secondary IC component from an unresolved population of MSPs using Fermi-LAT data alone. However, as discussed above, there is now growing evidence for a significant departure from spherical symmetry.

Nonspherical source morphologies have been explored on small ($<$1 kpc) scales. For example, in a region overlapping with the NB called the Galactic ridge, Ref.~\cite{Macias:2014sta} showed that several different CR scenarios can explain the multiwavelength data taken from this patch of the sky. In addition, Ref.~\cite{Carlson:2015ona} evaluated the impact of star-forming activity in the Galactic ridge on the GCE properties. The study by Ref.~\cite{Gaggero:2017jts} solved the diffusion equation for CRs with a position-dependent diffusion coefficient and explained the recent H.E.S.S.~\cite{Abramowski:2016mir} measurements from this region by the interaction of the CRs with the gas in the central molecular zone (CMZ). Reference~\cite{Guepin:2018jkb} explained the H.E.S.S. measurements by MSPs accelerating CR protons.

In this paper, we revisit the IC emission from MSP $e^\pm$ focusing on a potential \emph{nonspherical} source morphology related to the Galactic bulge structure. Reference~\cite{Abazajian:2014hsa} looked for and found a gamma-ray component following the IR distribution in the GC. The authors interpreted this as IC emission correlating with the distribution of optical light, but a propagation of the underlying $e^\pm$ was not performed. Here, we use the publicly available propagation code \texttt{GALPROP}\cite{Strong:1998fr,Galprop,Galpropsupplementary} and make detailed predictions for the spectrum and morphology of the IC emission at 100 MeV-100 TeV energies. For the spatial morphology of MSP $e^\pm$, we use a three-dimensional (3D) stellar distribution model for the Galactic bulge obtained from a fit to IR data~\cite{Freudenreich:1997bx} as well as one for the NB~\cite{Launhardt:2002tx} stars. We also show detailed comparisons against the spectrum and morphology obtained when the unresolved populations of MSPs are assumed to be spherically distributed. We find salient recognizable differences on large spatial scales and energies above $\sim$TeV, which can be used to distinguish source spatial morphologies.

This paper is structured as follows. In Sec.~\ref{sec:spatial} we describe the 3D models used for the spatial distribution of the putative MSP population in the GC. In Sec.~\ref{sec:prop} we provide details about the propagation setup and configuration of our \texttt{GALPROP} runs. Details about the assumed MSP injection spectra are also given in this section. Our main results are shown in Sec.~\ref{sec:results}, where we show the predicted spectrum and morphology for the IC emission at $\sim$GeV-TeV energies. We illustrate how future measurements of diffuse gamma-ray emission with the Cherenkov Telescope Array (CTA)~\cite{Wagner:2009cs,Consortium:2010bc} have the potential to constrain the main properties of this purported MSP population at the GC. Finally, we conclude our study in Sec.~\ref{sec:conclu}.

\section{Spatial distributions}\label{sec:spatial}

To propagate the $e^\pm$ injected by MSPs, we need to model their spectrum and the spatial distribution of MSPs. We first discuss the spatial distribution. Two scenarios are considered: the stellar mass distribution in the Galactic bulge, and the spherically symmetric model for comparison.

\subsection{Stellar models}\label{sec:stellar}

We describe our first scenario, in which an unresolved population of MSPs is created \emph{in situ} in the inner Galaxy and follows the distribution of stellar mass. Assuming that the same stellar populations responsible for the IR bulge trace the distribution of MSPs, a reasonable starting point for the MSP distribution is the bulge morphology itself. In principle, a proper morphological calculation would need to take into account the kick velocities of the MSP seeds at birth~\cite{Eckner:2017oul}. However, the kicks experienced by MSPs should be lower than for isolated pulsars, which is also necessary for them to be confined to globular clusters. For example, while isolated pulsars are consistent with a Maxwellian velocity distribution with dispersion of 190 km/s~\cite{Hansen:1997zw}, MSP estimates fall in the ranges $10-50$~\cite{Hooper:2013nhl,Cordes:1997my}, $85\pm13$~\cite{hobbs_manchester_teoh_hobbs_2004}, or at the high end $130\pm30$ km/s~\cite{Lyne1998kdn}. Thus for our MSP estimates we do not consider the effects of initial kicks, and we consider the spatial distribution of stars using the 3D Galactic bulge model of Ref.~\cite{Freudenreich:1997bx} and the NB model of Ref.~\cite{Launhardt:2002tx}.

\subsubsection{Galactic bulge}\label{sec:bulge}

The mid- and near-IR signals of the inner Galaxy stars reveal a bar structure. The Galactic bar makes a tilt angle $\theta_0$ from the GC-Sun direction while the Sun is located at $\sim$10 pc above the Galactic plane. Here we adopt the bar model derived from data taken with the Diffuse Infrared Background Experiment instrument on board the Cosmic Background Explorer~\cite{Freudenreich:1997bx}. The shape of the bar is a generalized ellipsoid which can be parametrized as
\begin{align}\label{eq:Rs}
  R_{\perp}^{C_{\perp}} &= \left(\dfrac{|X'|}{a_x}\right)^{C_{\perp}} + \left(\dfrac{|Y'|}{a_y}\right)^{C_{\perp}},\\
  R_s^{C_{\parallel}} &= R_{\perp}^{C_{\parallel}} + \left(\dfrac{|Z'|}{a_z}\right)^{C_{\parallel}},
\end{align}
where $R_s$ is the effective radius; $a_x$, $a_y$, and $a_z$ are the scale lengths; $C_\perp$ and $C_\parallel$ are the face-on and edge-on shape parameters; and $X'$, $Y'$, and $Z'$
are directions in the bar
coordinate. Reference~\cite{Freudenreich:1997bx} considered three different models for the radial dependence: model S, $\rho\propto\text{sech}^2(R_s)$; model E, $\rho\propto\exp(R_s^{-n})$; model P, $\rho\propto[1+(R_s/R_c)^n]$. We use model S here since it was the best-fit model found in Ref.~\cite{Freudenreich:1997bx}.

The models are truncated by a Gaussian function at the
radius $R_{\text{end}}$ with scale length $h_{\text{end}}$. For model S, the density of the bar~\footnote{We notice that there was a typo in the argument of the $\exp$ function in Eq.(14) of \cite{Freudenreich:1997bx} which has been corrected in our Eq.~\ref{eq:Rs}. We have confirmed this in private communication with H. Freudenreich.} is given by,
\begin{equation}\label{eq:rhobar}
  \rho_{\text{bar}}\propto \begin{cases}
    \text{sech}^2(R_s), & R \leq R_{\text{end}},\\
    \text{sech}^2(R_s)e^{-\frac{(R-R_{\text{end}})^2}{h_{\text{end}}^2}}, &   R > R_{\text{end}}.
  \end{cases}
\end{equation}

The bar parameters used in our work are displayed in Table~\ref{tab:modelS}. These correspond to the best-fit values for model S using the so-called primary mask~\cite{Freudenreich:1997bx}. Recent studies of the bulge suggest larger tilt angles of $\sim 30^\circ$~\cite{doi:10.1093/mnras/stt1045,Portail:2016vei}. However, we keep the best-fit angle from Ref.~\cite{Freudenreich:1997bx} since it is consistent with the ISRF implemented in \texttt{GALPROP} v54. Also, as we discuss later, we do not expect the IC emissions to be very sensitive to this angle. Overall, the stellar mass of the Galactic bulge is $(1.4-1.7)\times 10^{10}$ solar masses $(M_\odot)$~\cite{2016ARA&A..54..529B}.

\begin{table}[t!]
  \caption{\label{tab:modelS} The parameter values for the Galactic bar model S of Ref.~\cite{Freudenreich:1997bx}.}
  \begin{ruledtabular}
    \begin{tabular}{ll}
      Parameters & Model S \\ \hline
      Distance to the Galactic Plane $Z_0$ (pc) & 16.46 $\pm$ 0.18\\
      Bar Tilt Angle $\theta_0$ (deg) & 13.79 $\pm$ 0.09\\
      Bar $X$ Scale Length $a_x$ (kpc) & 1.696 $\pm$ 0.007 \\
      Bar $Y$ Scale Length $a_y$ (kpc) & 0.6426 $\pm$ 0.0020 \\
      Bar $Z$ Scale Length $a_z$ (kpc) & 0.4425 $\pm$ 0.0008 \\
      Bar Cutoff Radius $R_{\text{end}}$ (kpc) & 3.128 $\pm$ 0.014 \\
      Bar Cutoff Scale Length $h_{\text{end}}$ (kpc) & 0.461 $\pm$ 0.005 \\
      Bar Face-On Shape $C_\perp$ & 1.574 $\pm$ 0.014 \\
      Bar Edge-On Shape $C_\parallel$ & 3.501 $\pm$ 0.016
    \end{tabular}
  \end{ruledtabular}
\end{table}

\subsubsection{Nuclear bulge}\label{sec:nb}

The NB refers to a dense stellar structure contained in the innermost region of the Galaxy. Associated with the CMZ, the NB has younger stars and undergoes active star formation, distinguishing it from the old and evolved stars of the Galactic bulge~\cite{Launhardt:2002tx}. The NB makes up around 10\% of the stellar mass in the bulge and its gamma-ray luminosity is comparable with that of the Galactic bulge~\cite{Macias:2016nev,Bartels:2017vsx}. The NB resides within the inner 230 pc of the GC and is made of two components:

\paragraph{Nuclear stellar cluster (NSC):} The NSC is a relatively small and very dense spherically symmetric structure in the innermost part of the NB. The stellar density in this region has been shown~\cite{Launhardt:2002tx} to be well described by a simple radial power-law function
\begin{equation}\label{eq:NSC}
  \rho_{\text{NSC}}(R)=\dfrac{\rho_0}{1+\left(\dfrac{R}{R_0}\right)^{n}},
\end{equation}
with best-fit power-law indices $n = 2.0$ for $R \leq 6$ pc and $n = 3.0$ for $R > 6$ pc, with core radius fixed to $R_0 = 0.22$ pc. The stellar mass of the entire NSC is $(3\pm$ 1.5) $\times$ 10$^7$ $M_\odot$.

\paragraph{Nuclear stellar disk (NSD):} Surrounding the NSC is the NSD which makes up most of the stellar mass of the NB. The NSD is a cylindrical object with a radial dependence approximately described by a broken power-law function,
\begin{equation}\label{eq:rhonsd}
  \rho_{\text{NSD}}(r) = \begin{cases}
    \rho_0\ r^{-0.1}, & r < 120\ \text{pc},\\
    \rho_1\ r^{-3.5}, & 120\ \text{pc} \leq r < 220\ \text{pc},\\
    \rho_2\ r^{-10}, & r \geq 220\ \text{pc}.
  \end{cases}
\end{equation}
The scale densities $\rho_0$, $\rho_1$, and $\rho_2$ ensure the continuity of the NSD density function. The density variation along the $z$ direction is given by an exponential cutoff with a scale height 45 $\pm$ 5 pc. The stellar mass of the entire NSD is (1.4 $\pm$ 0.6) $\times$ 10$^9$ $M_\odot$.

\subsection{Spherically symmetric source}\label{sec:NFW}

Although recent reanalyses of the GCE~\cite{Macias:2016nev,Bartels:2017vsx} have shown that the Fermi-LAT data from the inner Galaxy prefer stellar maps to spherically symmetric ones, here for comparison purposes, we also model the putative MSP population at the GC with the square of an NFW density profile, of the form
\begin{eqnarray}\label{eq:NFW}
  \rho(R)_{\text{NFW}} = \dfrac{\rho_0}{\left(\dfrac{R}{R_\odot}\right)^\gamma\left(\dfrac{1+R/R_s}{1+R_s/R_\odot}\right)^{(3-\gamma)}},
\end{eqnarray}
where we use a core radius $R_s$ = 23.1 kpc, Sun-GC distance $R_\odot$ = 8.25 kpc, and an inner slope of $\gamma$ = 1.20~\cite{Abazajian:2012pn,Macias:2013vya}. We label this as NFW$^2$.

\section{Propagation}\label{sec:prop}

\subsection{\texttt{GALPROP} code}\label{sec:galprop}

We used the publicly available software package \texttt{GALPROP} v54~\cite{Galprop,Galpropsupplementary} in order to calculate the secondary gamma-ray emission from CR $e^\pm$ injected by MSPs. GALPROP is a numerical tool that solves the particle transport equations for a given source distribution and boundary conditions for all species of CRs. In particular, CRs can get accelerated by a multitude of different sources and then propagate long distances in the Galaxy. During propagation they produce secondary particles via interactions with the ISM and ISRF. Galactic diffuse gamma-ray emission is produced via $\pi_0$ decay, bremsstrahlung, and IC scattering, while lower-energy emission is produced via synchrotron radiation.

The CR $e^\pm$ transport equation is a partial differential equation of the form
\begin{eqnarray}
  \label{eq:prop_eq}
  \dfrac{d\psi}{dt} =\ &&q(\vec{r},p)+\vec{\nabla}\cdot(D_{xx}\vec{\nabla}\psi-\vec{V}\psi)\nonumber\\
                       &&+\dfrac{\partial}{\partial p}p^2D_{pp}\dfrac{\partial}{\partial p}\dfrac{1}{p^2}\psi-\dfrac{\partial}{\partial p}[\dot{p}\psi-\dfrac{p}{3}(\vec{\nabla}\cdot\vec{V})\psi],
\end{eqnarray}
where $\psi = \psi(\vec{r},p,t)$ is the density per unit of the total particle momentum, $D_{xx} = \beta D_0 (\frac{\rho}{\rho_0})^\delta$ is the spatial diffusion coefficient where $\rho$ is the rigidity of the $e^\pm$ and $\beta = v/c$, $\vec{V}$ is the convection velocity and $q(\vec{r},r)$ is the source term. Reacceleration is introduced as diffusion in momentum space with coefficient $D_{pp}$, which is related to $D_{xx}$ and the Alfv\'{e}n speed $v_A$. We refer interested readers to the excellent review in Ref.~\cite{Strong:2007nh} for more details. We keep convection off, $\vec{V}=0$.

Given the source function of MSPs, the IC energy losses at each spatial bin are calculated by
\begin{equation}
  \dot{p} = \int d(\log{\nu})\dfrac{\nu U_\nu}{\hbar \nu}\dfrac{dp}{dt}(\nu,\gamma),
\end{equation}
where $\nu$ and $U_\nu$ are the frequency and energy density of the ISRF, and $\gamma$ is the $e^\pm$ Lorentz factor. Very high-energy photons can interact with the photon fields and pair produce additional $e^\pm$ ($\gamma\gamma \to e^+e^-$). This is not included in \texttt{GALPROP} v54; however it only becomes important at $\sim$100 TeV, where the survival probability from the Galactic center to Earth reaches a minimum of 75-80\%~\cite{Moskalenko:2005ng}. At 10 TeV, the effect is reduced to the percent level.

GALPROP v54 contains dedicated routines to compute the propagation of DM annihilation/decay products and predict sky maps of secondary emissions. We modify the \texttt{gen\_DM\_source.cc} routine, which allows for user-defined source functions of the DM yields (DM profile and particle spectra), to model  $e^\pm$ injected from MSPs. To compute the IC sky maps, we turn off the propagation of non-MSP CRs. As a first step, we made detailed comparisons of the results obtained with our modified GALPROP package against the literature~\cite{Cirelli:2014lwa,Cirelli:2016mrc,Yuan:2014yda,Petrovic:2014xra}. We confirm that we are able to reproduce the gamma-ray spectrum and spatial profiles of Ref.~\cite{Cirelli:2014lwa} as well as the synchrotron sky maps given in Ref.~\cite{Cirelli:2016mrc} in the context of DM annihilations. Of greater relevance to this study are our detailed checks of the results in Refs.~\cite{Yuan:2014yda, Petrovic:2014xra}. Although we were able to reproduce the gamma-ray spectral and spatial profiles in Ref.~\cite{Petrovic:2014xra}, we were only able to obtain the spectra given by Ref.~\cite{Yuan:2014yda}. There are differences between our predicted spatial maps and those given in Fig.~3 of Ref.~\cite{Yuan:2014yda} for the same propagation setup. However, we believe these differences could be due to their using older two-dimensional (2D) ISRF maps in GALPROP.

\subsection{Source function of MSP \texorpdfstring{$e^\pm$}{electrons/positrons}}\label{sec:source}

The spin-down energy $\dot E$ of MSPs is responsible for generating relativistic $e^\pm$ winds. The accelerations of these $e^\pm$ are limited when they lose energy via curvature radiation in the pulsar magnetosphere. Gamma rays are generated in this process. The gamma-ray efficiency $L_\gamma/\dot E$ is estimated to be about 10\% on average \cite{TheFermi-LAT:2013ssa}. The $e^\pm$ that escape the magnetosphere via open field lines carry a fraction $f_{e^\pm}$ of the spin-down energy into the interstellar environment. The $e^\pm$ injection luminosity of MSPs is therefore related to the gamma-ray luminosity by
\begin{equation}\label{eq:Le}
  L_{e^\pm}=f_{e^\pm}\dot{E}=10f_{e^\pm}L_\gamma.
\end{equation}
The High-Altitude Water Cherenkov Experiment (HAWC) observations of Geminga and PSR B0656+14 in the TeV energy range suggest $f_{e^\pm}$ values of $\sim$7.2-29\%~\cite{Hooper:2017gtd}. Constraints by the H.E.S.S.~observations also indicate $f_{e^\pm}$ of the order 10\% if about a thousand MSPs reside in the nuclear stellar cluster around Sgr A$^*$~\cite{Bednarek:2013oha}.

We normalize the two spatial templates we implement in \texttt{GALPROP} (stellar and NFW$^2$) by Eq.~(\ref{eq:Le}), using the best-fit gamma-ray luminosities obtained in Ref.~\cite{Macias:2016nev}. The gamma-ray luminosities for the stellar template (which is a linear combination of the Galactic bar and NB models; see Sec.~\ref{sec:stellar}) are $L^{\text{bar}}_\gamma$ = (1.4 $\pm$ 0.2) $\times$ 10$^{37}$ and $L^{\text{NB}}_\gamma$ = (4.0 $\pm$ 1.0) $\times$ 10$^{36}$ erg s$^{-1}$ for the bar and NB, respectively, while the NFW$^2$ template (see~\ref{sec:NFW}) has $L^{\text{NFW}^2}_\gamma$ = (1.7 $\pm$ 0.2) $\times$ 10$^{37}$ erg s$^{-1}$. These luminosities were taken from an analysis region of size 15$^\circ$ $\times$ 15$^\circ$ around the GC~\cite{Macias:2016nev}.

The source term $q(\vec{r},E)$ (in units of  MeV$^{-1}$ cm$^{-2}$ s$^{-2}$ sr$^{-1}$) that is included in \texttt{GALPROP} can be written as the product of the injection spectrum $dN/dEdt$ and the source density distribution $\rho(\vec{r})$,
\begin{equation}
  q(\vec{r},E) = \frac{c}{4\pi} N_0 \dfrac{dN}{dEdt}\rho(\vec{r}).
\end{equation}
The factor $c/4\pi$ is a convention in the {\tt GALPROP} code. The source function is normalized by $N_0$, such that the integration over energy and volume matches Eq.~(\ref{eq:Le}),
\begin{equation}
  \label{eq:norm}
  N_0 \int E\dfrac{dN}{dEdt}dE \int\rho(\vec{r})dr^3 = L_{e^\pm} = 10 f_{e^\pm} L_\gamma.
\end{equation}
We will explore a range of $e^\pm$ spectra as detailed in Sec.~\ref{sec:spectrum}.

\begin{figure}[t!]
  \includegraphics[width = \columnwidth]{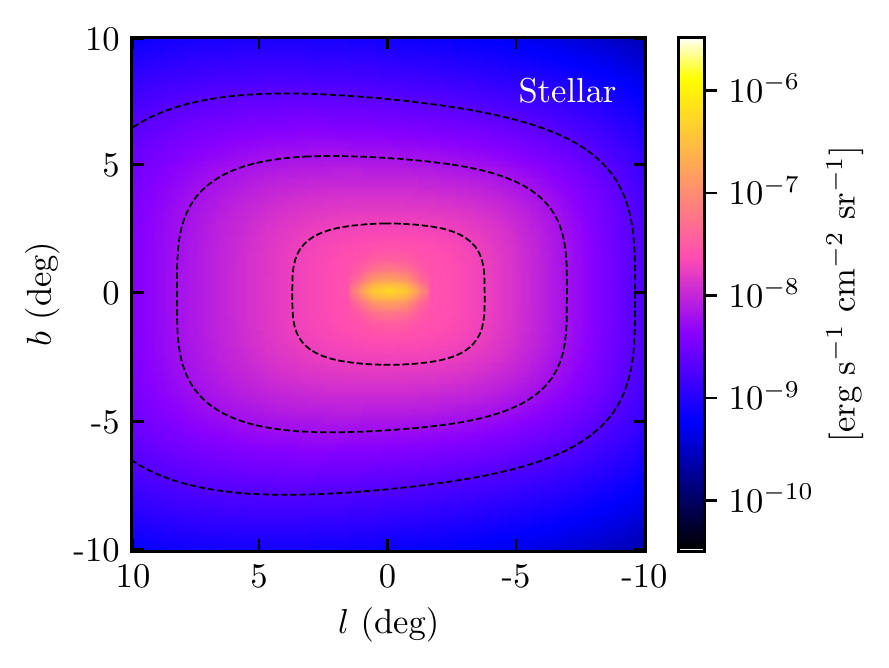}
  \includegraphics[width = \columnwidth]{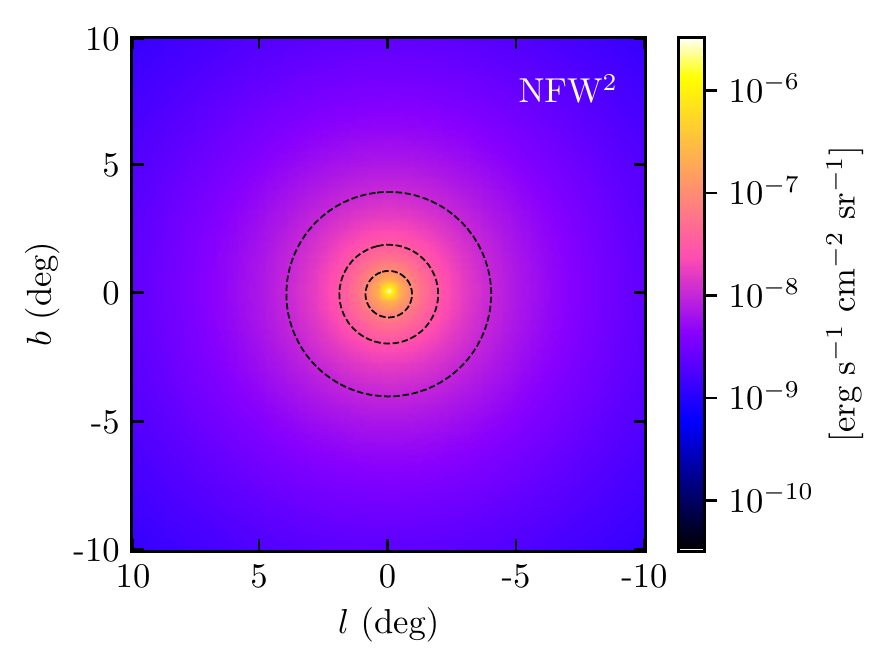}
  \caption{MSP $e^\pm$ source distributions for the stellar (top panel) and the NFW$^2$ (bottom panel) template over a $20^\circ \times 20^\circ$ region around the GC. The normalizations are determined by gamma-ray luminosities through Eq.~(\ref{eq:Le}) (see text).
The source distributions are noticeably different. While the stellar template is rectangular and asymmetric due to the tilting angle between the Galactic bar's long axis and the GC-Sun direction, the NFW$^2$ template is spherically symmetric. The very bright quasielliptical region in the central few degrees in the top panel corresponds to the NB stellar component.}
  \label{fig:injection_skymap}
\end{figure}
\begin{figure}[t!]
  \includegraphics[width = \columnwidth]{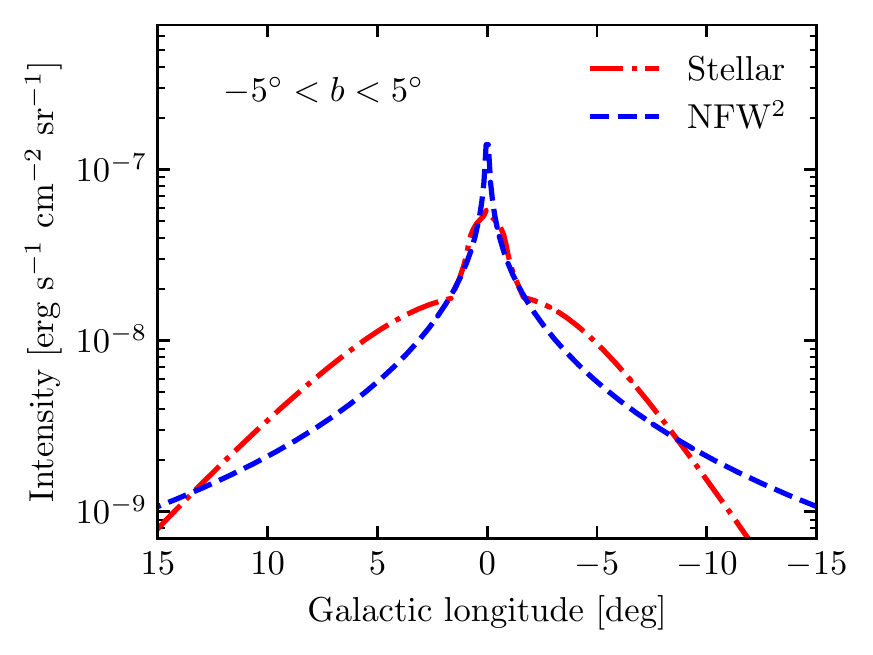}
  \includegraphics[width = \columnwidth]{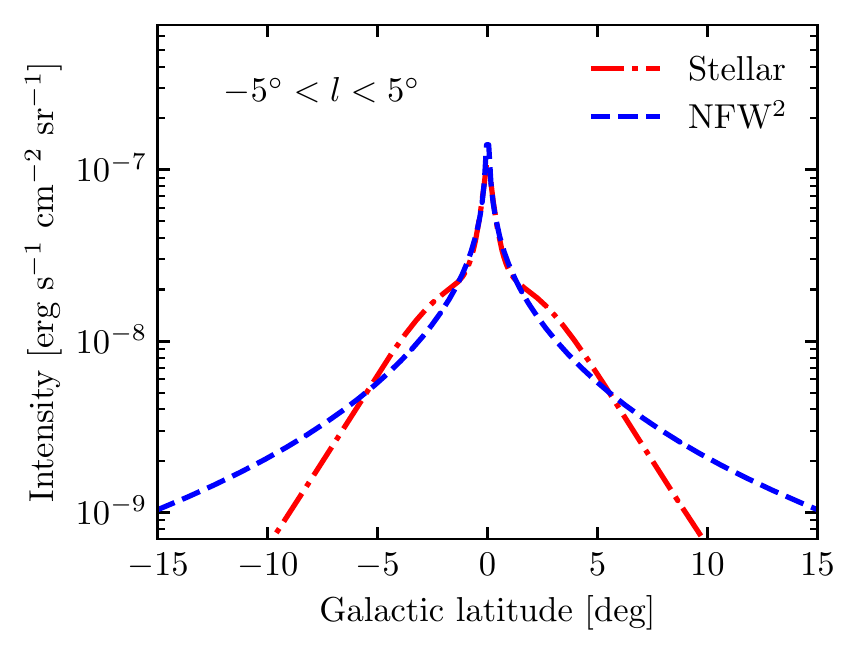}
  \caption{Longitudinal (top panel) and latitudinal (bottom panel)
    profiles for the MSP $e^\pm$ source distributions considered in this work. The profiles
    are for 10$^\circ$ wide bands and show the longitude and latitude
    from $15^\circ$ to $-15^\circ$. Noticeable morphological differences are seen in both directions.}
  \label{fig:injection_profile}
\end{figure}

To show the source distribution and their relative injection intensities we produce injection intensity maps obtained by integrating along the line-of-sight direction the corresponding source density models,
\begin{equation}
  \label{eq:j_factor}
  I(l,b) \propto \int_{\text{l.o.s}} \rho(s,l,b) ds,
\end{equation}
where $l$ and $b$ are the galactic longitude and latitude and $s$ is the distance from the Sun along the line-of-sight direction.
Figure~\ref{fig:injection_skymap} (top panel) displays the injection intensities of the $\mbox{Galactic bar}+\mbox{NB}$ template in a 20$^\circ$ $\times$ 20$^\circ$ region around the GC. The resulting sky map is oblate and asymmetric. It is brighter for positive longitudes since the long axis of the bar makes a tilting angle with the GC-Sun direction. In contrast, Fig.~\ref{fig:injection_skymap} (bottom panel) shows that the NFW$^2$ source is spherically symmetric and has a strong peak in the center of the Galaxy. We also present corresponding longitudinal and latitudinal profiles for the two models in Fig.~\ref{fig:injection_profile}. The aforementioned longitudinal asymmetry in the stellar distribution can clearly be seen in the top panel. Here it is also noticeable that the NFW$^2$ intensity profile is more strongly peaked than the stellar template. The oblateness of the $\mbox{Galactic bar}+\mbox{NB}$ model is made manifest by comparing the tails of the profiles in both panels.

\subsection{Injection spectrum}\label{sec:spectrum}

The maximum energy that MSPs can accelerate $e^{\pm}$ to is estimated to be $\sim 100$ TeV. In this work we adopt a similar $e^\pm$ spectrum to that used in Ref.~\cite{Yuan:2014yda} and explore a range of parameter values. Namely, we use a power law with an exponential cutoff,
\begin{equation}
  \label{eq:msp_spectrum}
  \dfrac{dN}{dEdt} \propto E^{-\Gamma}\exp(-E/E_{\text{cut}}),
\end{equation}
where $\Gamma$ is the spectral slope and $E_{\text{cut}}$ is the energy cutoff. We vary $\Gamma$ between 1.5 and 2.5, while $E_{\text{cut}}$ is varied between 10 and 100 TeV. The combinations of spectral parameters assumed in this work are listed in Table~\ref{tab:msp_spectrum}. Note that our choice of spectral parameters represents the entire MSP population in the Galactic bulge.
\begin{table}[t!]
  \centering
  \caption{Injection spectral parameters of MSP $e^{\pm}$ adopted in this work. See Eq.~(\ref{eq:msp_spectrum}).}
    \begin{tabular}{lcc}\hline\hline
    Model Name&$\Gamma$ & $E_{\text{cut}}$\\
    & &  (TeV) \\\hline
    Baseline &2.0 & 50 \\
    Inj1&1.5 & 50 \\
    Inj2&2.5 & 50 \\
    Inj3&2.0 & 10 \\
    Inj4&2.0 & 100\\\hline\hline
    \end{tabular}
  \label{tab:msp_spectrum}
\end{table}

\subsection{Configurations}

In order to make simulations of CR propagation that are as realistic as possible, it is crucial to understand the diffusion parameters for all the CR species. Reference~\cite{Johannesson:2016rlh} performed a scan of the parameter space of the CR injection and propagation. The scan was done separately for the low-mass isotopes ($p$, $\bar{p}$ and He) and the light elements (Be, B, C, N, O). Since each set of species has a different lifetime they probe different regions of the Galaxy. They found that the best-fit parameter setup for the low mass isotopes is different from that obtained for the heavier elements. Here, we adopt the propagation parameters for the low-mass isotopes of Ref.~\cite{Johannesson:2016rlh}. We account for uncertainties in the propagation parameters by also including the 95\% credible contours provided in their 2D marginalized posterior distributions. The propagation setups used in this work are listed in Table~\ref{tab:dpara}.

We perform 3D \texttt{GALPROP} simulations using the standard ISRF data available with version 54 of the software package. The calculations are made for a Cartesian spatial grid with the Galactic plane placed in the $X$-$Y$ plane and the GC located at the origin of the coordinate system. The $X$ axis is defined by the GC-Sun direction.

In our simulations, the propagation volume extends to $\pm$20 kpc in both the $X$ and $Y$ direction. However, the CR halo height $z_h$ is set to different values depending on the model considered, as shown in Table~\ref{tab:dpara}. To trace the physics of the NB, we performed resolution tests for the spatial grid sizes. We fixed $\Delta Z$ = 0.125 kpc and found that the IC spectra and sky maps showed little difference when the resolution was $\Delta X$ = $\Delta Y$ = 0.125 or 0.25 kpc. We therefore adopted $\Delta X$ = $\Delta Y$ = $\Delta Z$ = 0.125 kpc for all our runs except for the $z_h = 19.28$ kpc run, which was performed at a lower resolution of $\Delta X$ = $\Delta Y$ = 0.25 and $\Delta Z$ = 0.125 kpc due to computing memory demands. As a result, each run takes about 80 hours to finish using a computer cluster node with 500 GB memory and running at $\sim 6 \times 10^{11}$ flop/s.

After choosing the resolution, Eq.~(\ref{eq:rhobar}) can be converted to the number of MSPs per spatial bin and normalized to reproduce the total number of MSPs that inhabit the Galactic bulge~($\sim 10^4$ \cite{Gonthier:2018ymi}). We find that every (0.125 kpc)$^3$ bin corresponds to $\sim 40$ MSPs at the GC, and $\sim 4$ MSPs at 3 kpc along the long axis of the Galactic bar. We thus approximate the MSP distribution as a smooth function in our simulations.

\begin{table}[t!]
  \centering
  \caption{\label{tab:dpara} Propagation parameter setups considered in this study. Our baseline model corresponds to the best-fit propagation parameter in the scan for low-mass isotopes in Ref.~\cite{Johannesson:2016rlh}, while the five additional models reflect the 95\% confidence contours in their propagation parameter scan \cite{Johannesson:2016rlh} (see text).}
  \begin{ruledtabular}
    \begin{tabular}{ c c c c c }
      & $D_0$ & $z_h$ & $v_A$  &$\delta$ \\
      & (10$^{28}$ cm$^2$ s$^{-1}$) & (kpc) & (km s$^{-1}$) & \\ \hline
      Baseline & 6.330 & 9.507 & 8.922 & 0.466 \\
      Model 1 & 3.159 & 9.507 & 8.922 & 0.466 \\
      Model 2 & 7.006 & 9.507 & 8.922 & 0.573 \\
      Model 3 & 8.072 & 9.507 & 8.922 & 0.351 \\
      Model 4 & 2.748 & 3.000 & 8.922 & 0.466 \\
      Model 5 & 7.742 & 19.280 & 8.922 & 0.466 \\
    \end{tabular}
  \end{ruledtabular}
\end{table}

We adopt the best-fit propagation of Ref.~\cite{Johannesson:2016rlh} with the default $e^\pm$ spectrum of Ref.~\cite{Yuan:2014yda} as our baseline setup. In order to evaluate the impact of different propagation and spectral assumptions, we consider different propagation model setups in Table~\ref{tab:dpara} and the different $e^{\pm}$ injection spectra listed in Table~\ref{tab:msp_spectrum}. For our baseline spectral setup we explore all variations in propagation setup. For our baseline propagation setup we explore all $e^{\pm}$ injection spectral combinations. In all our simulations the efficiency of the MSP $e^\pm$ is fixed to $f_e^\pm$ = 0.1.

\subsection{Magnetic field}\label{sec:bfield}

We adopt the default magnetic field from \texttt{GALPROP}, which is a double-exponential function,
\begin{equation}
	B(r,z) = B_0 \exp{\left(-\dfrac{r-R_\odot}{R_0}\right)}\exp{\left(-\dfrac{z}{z_0}\right)},
\end{equation}
where $B_0 = 5\ \mu$G is the local magnetic field at the Solar System radius, and the scale parameters $R_0$ = 10 kpc, and $z_0$ = 2 kpc. The magnetic field strength of this model matches the 408 MHz synchrotron data~\cite{Strong:1998fr} and is in agreement with the total Galactic magnetic field estimates in the literature~\cite{1995ASPC...80..507H,2001SSRv...99..243B}. However, the magnetic field at the center of the Galaxy remains uncertain. In particular, a multiband modeling on scales of 400 pc about the GC has produced a lower limit of 50 $\mu$G on the magnetic field strength \cite{Crocker:2010xc}, which the default \texttt{GALPROP} magnetic field does not obey (yielding, instead, a field strength of $\sim 10\ \mu$G). To this end, we test a modified magnetic field where we set $B = 50\ \mu$G within a 400 pc region around the GC, but otherwise it matches the \texttt{GALPROP} default field everywhere else. The impacts of such a magnetic field will be discussed.

\begin{figure}[t!]
  \includegraphics[width=\columnwidth]{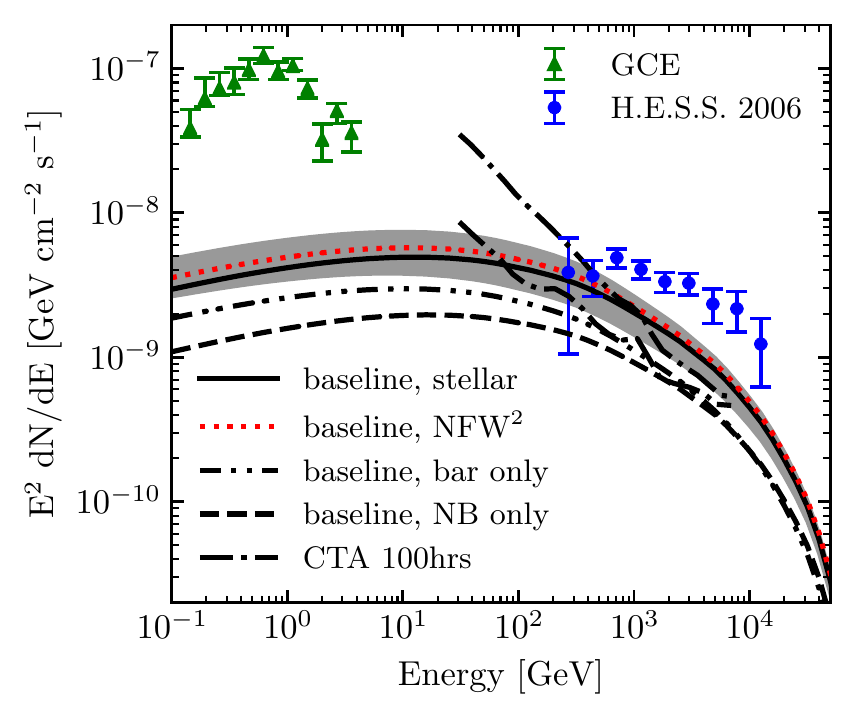}
  \caption{IC emissions from the Galactic ridge ($6^\circ\times 2^\circ$ area around the GC). The fluxes from the baseline model for the stellar template (black solid) and NFW$^2$ template (red dotted) are shown. The shaded band represents the uncertainties due to the propagation parameters for the stellar template. The NB (dashed) and the Galactic bar (dot-dot-dashed) are shown separately for the baseline stellar model. Green triangles show the GCE data~\cite{Macias:2016nev} and blue circles show the H.E.S.S. residuals~\cite{Aharonian:2006au} in the Galactic ridge region. The upper dot-dashed (lower dot-dashed) line shows the CTA sensitivity for 100 hours from an ON-OFF analysis using the ring method toward the GC with (without) systematic uncertainty considerations~\cite{Silverwood:2014yza}. A dedicated morphological analysis could improve the shown sensitivities by up to a factor of $\sim 10$ \cite{Silverwood:2014yza}.}
  \label{fig:spectrum_diffuse}
\end{figure}

\section{\label{sec:results} Results}

Figure~\ref{fig:spectrum_diffuse} shows the predicted IC emissions from the Galactic ridge region (defined as the $6^\circ\times 2^\circ$ area around the GC) for our stellar template (Galactic bar + NB). The solid black line is the expected flux from the baseline model. The shaded band represents the uncertainties resulting from changing the propagation setups as listed in Table~\ref{tab:dpara}. In the same plot we also show the expected flux from the NFW$^2$ template (red dotted), which is indistinguishable from that from the stellar template given the uncertainties from the propagation parameters. This conclusion also holds for larger regions of the bulge of interest ($20^\circ\times 20^\circ$ around the GC). For the stellar template, we also show the IC emissions from the NB and the Galactic bar separately. For the Galactic ridge, the NB contributes $\sim 1/3$ of the total IC emission in the GeV energy range, and $\sim 1/2$ in the TeV range.

\begin{figure}[t!]
  \includegraphics[width=\columnwidth]{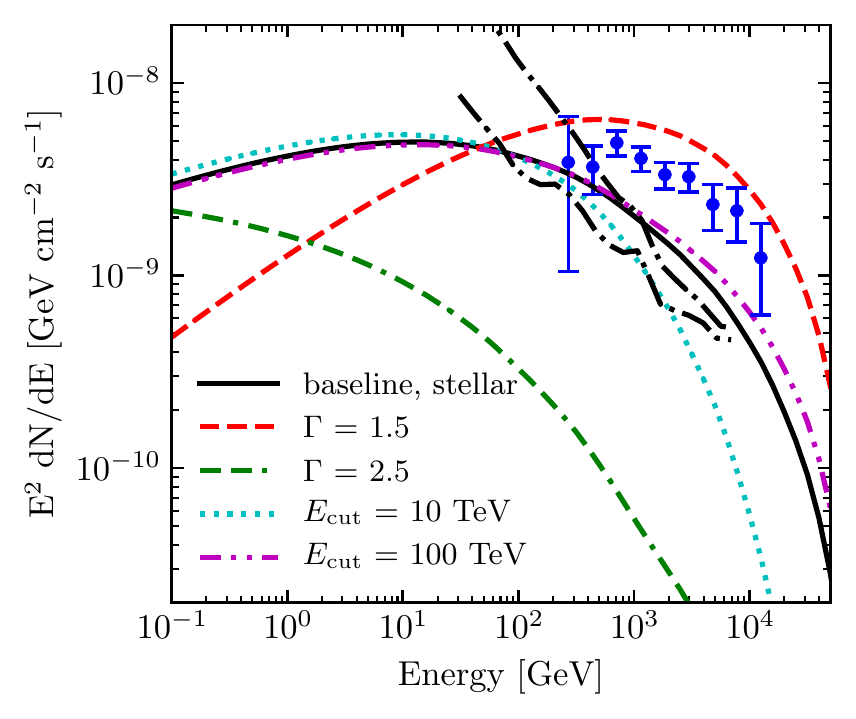}
  \caption{IC emissions for different $e^\pm$ spectra as listed in Table~\ref{tab:msp_spectrum}. H.E.S.S.~data and CTA sensitivities are shown in the same way as in Fig.~\ref{fig:spectrum_diffuse}, but note the different $y$ axis range.  Most spectra are below the H.E.S.S. measurements except for the hardest spectrum $\Gamma$ = 1.5 with $E_{\text{cut}}$ = 50 TeV which could be alleviated by a lower fraction of spin-down energy into relativistic $e^\pm$. For the soft spectrum $\Gamma$ = 2.5, the fluxes are below the CTA sensitivities by more than an order of magnitude.}
  \label{fig:spectrum_spectral}
\end{figure}

The GCE data~\cite{Macias:2014sta} at GeV energies, scaled to the Galactic ridge region, are shown by green triangles. The IC emission from MSPs is predicted to contribute less than $\sim 10$ \% of the GCE. This is consistent with previous works that have not found evidence for secondary emission at the $\sim 1$ GeV energy range~\cite{Lacroix:2015wfx}. The IC fluxes extend to the TeV energy range before decreasing steeply at $\sim 10$ TeV. Meanwhile the predicted fluxes are below or within the range of H.E.S.S.~observations (blue circles)~\cite{Aharonian:2006au}. Our resolution does not probe the smaller-scales of $0.2^\circ$--$0.5^\circ$ investigated in Ref.~\cite{Hooper:2018fih}. Also shown in the same figure are the differential sensitivities for 100 hours of CTA diffuse GC observations~\cite{Silverwood:2014yza} (dot-dashed black lines). We note that our predicted IC fluxes could be detected by forthcoming CTA observations. Reference~\cite{Silverwood:2014yza} computed the CTA sensitivities to a DM annihilation signal in the GC. Although the spatial morphology considered in their work is different than ours, we use it as a representative of the CTA sensitivities to a large diffuse signal in this sky region. That work also showed that by performing dedicated morphological analyses, the sensitivity of CTA can be improved by up to an order of magnitude for DM annihilation signals. We expect that similar improvements could be obtained by performing such morphological analyses with the models considered in our current study.

\begin{figure*}[t!]
	\includegraphics[width=0.68\columnwidth]{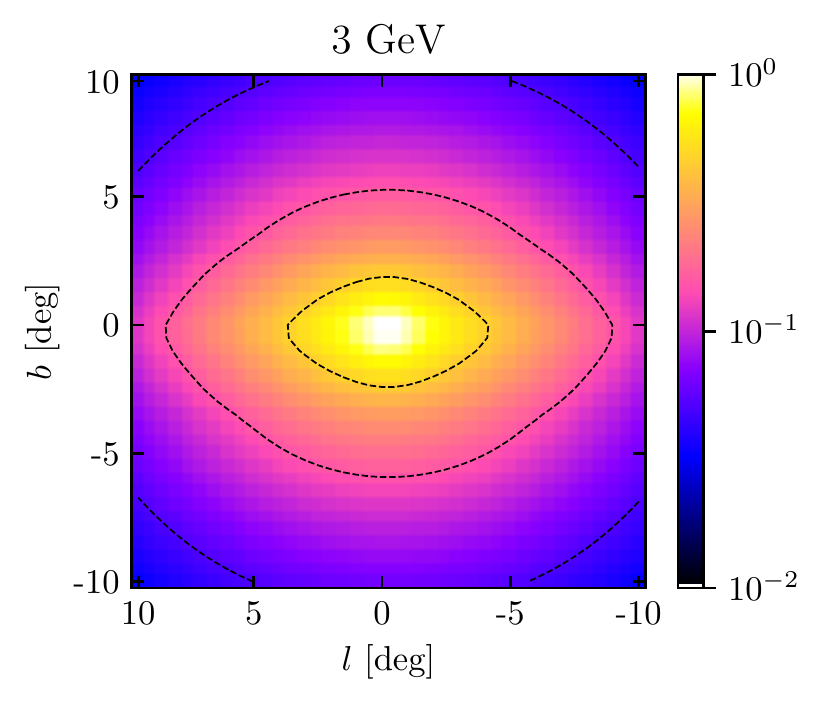}
    \includegraphics[width=0.68\columnwidth]{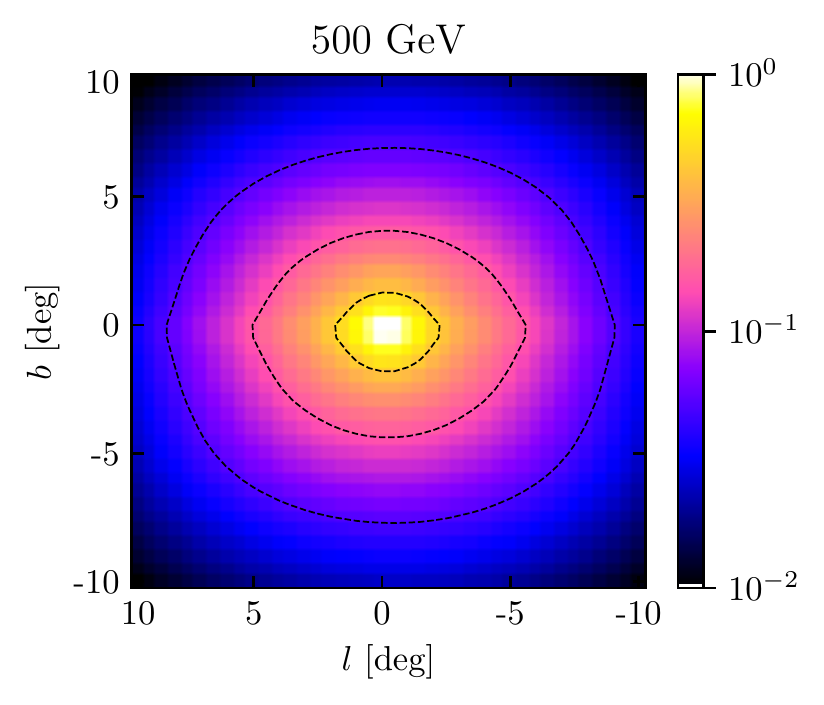}
     \includegraphics[width=0.68\columnwidth]{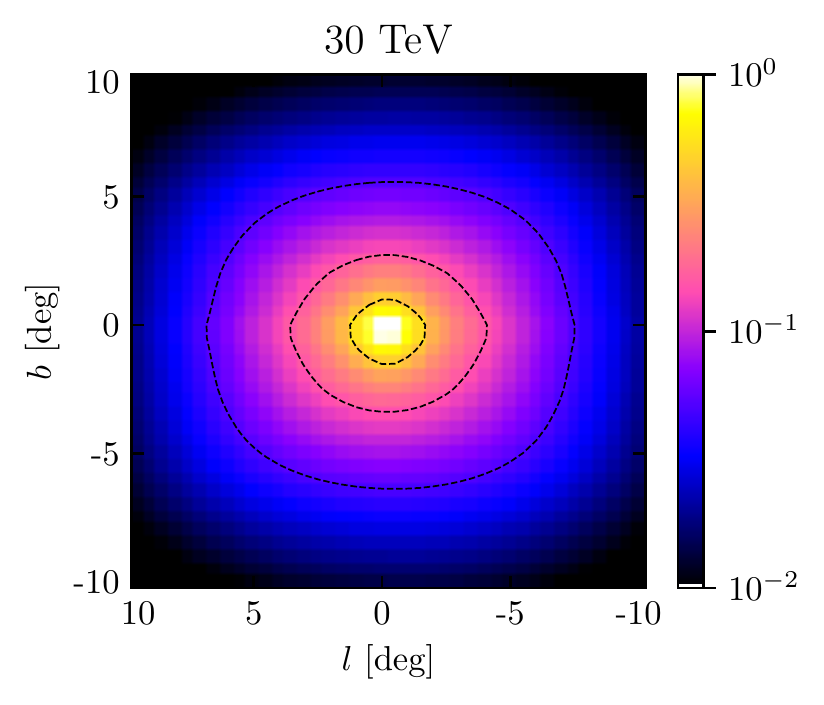}
    \includegraphics[width=0.68\columnwidth]{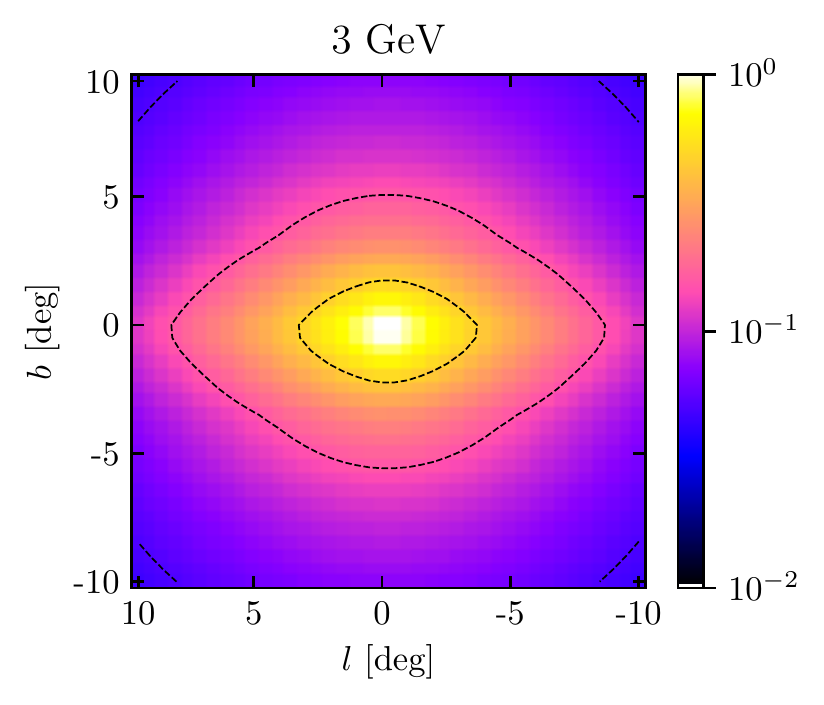}
    \includegraphics[width=0.68\columnwidth]{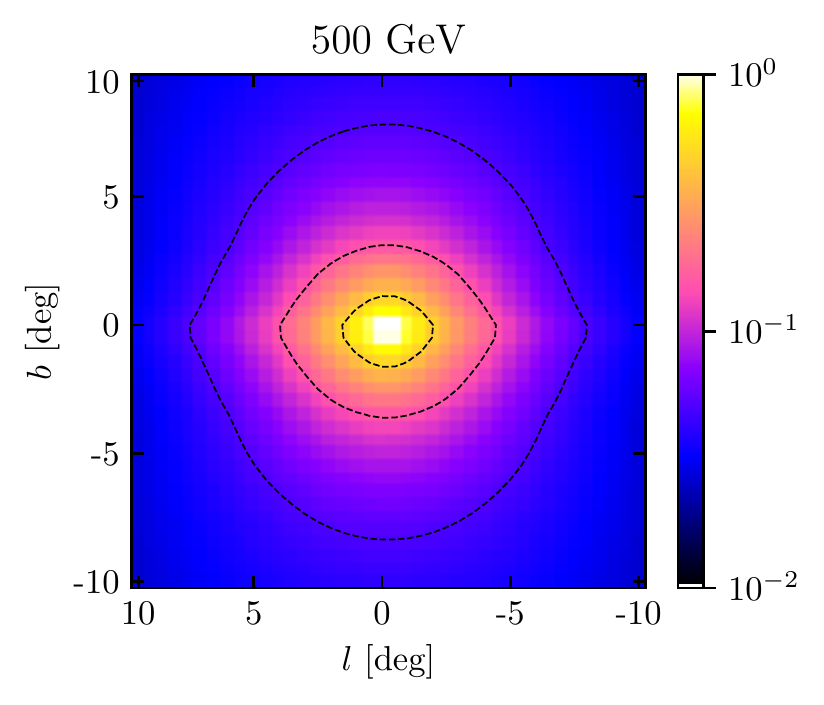}
    \includegraphics[width=0.68\columnwidth]{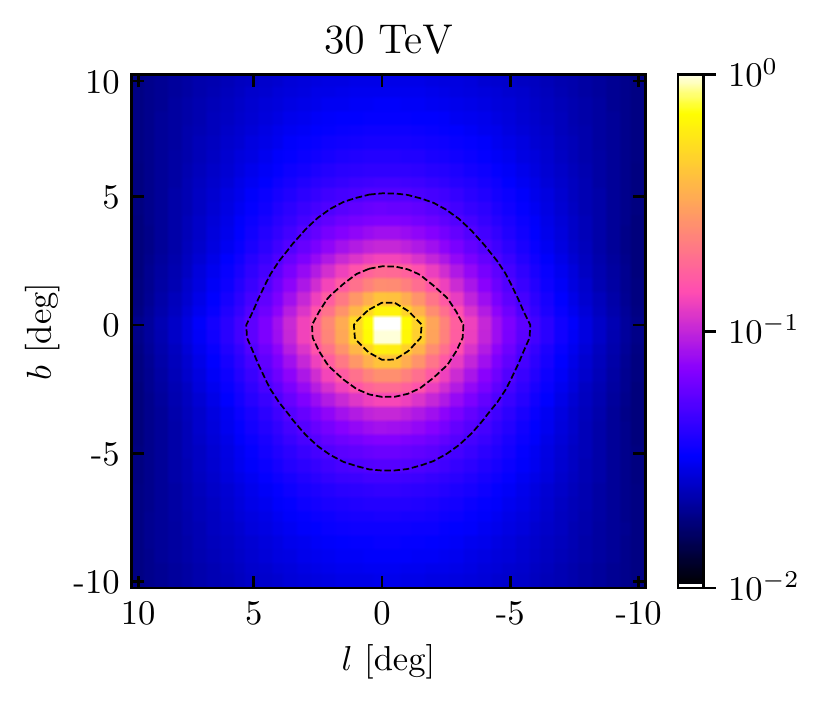}
  \caption{IC sky maps from the baseline model for the stellar template (top row) and the NFW$^2$ template (bottom row) over a $20^\circ\times 20^\circ$ region around the GC. Three energy windows are shown: 3 GeV (left column), 500 GeV (center column), and 30 TeV (right column). The injected $e^\pm$ are normalized by gamma-ray luminosities. The 5\%, 15\%, and 45\% flux levels with respect to the GC are shown by dotted contours. The morphological differences between the stellar template and the NFW$^2$ template appear more prominently at high energies.}
  \label{fig:skymap_bulge_vs_nfw}
	\includegraphics[width=0.68\columnwidth]{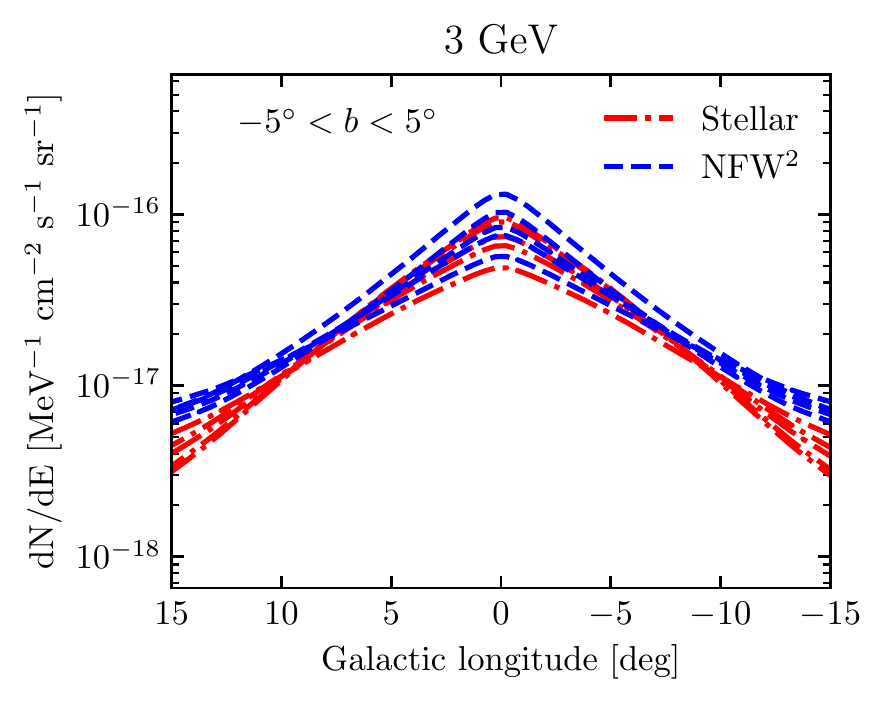}
    \includegraphics[width=0.68\columnwidth]{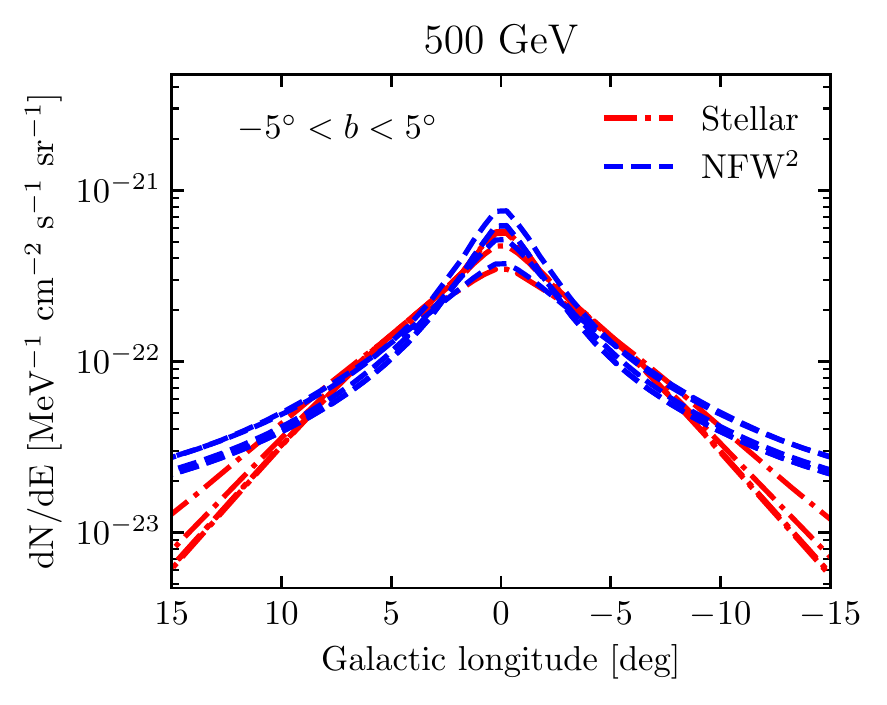}
	\includegraphics[width=0.68\columnwidth]{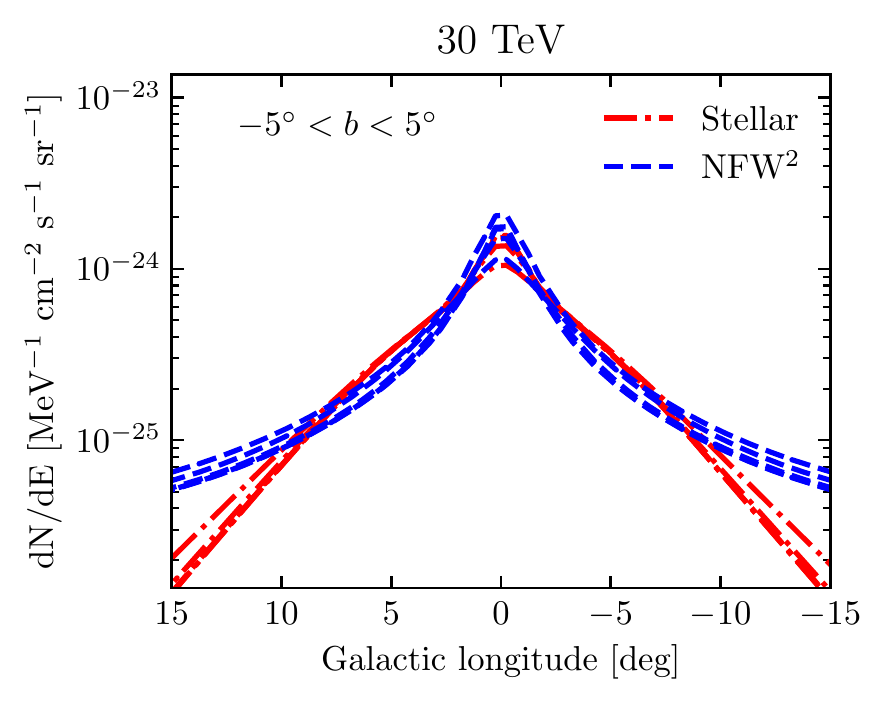}
	\includegraphics[width=0.68\columnwidth]{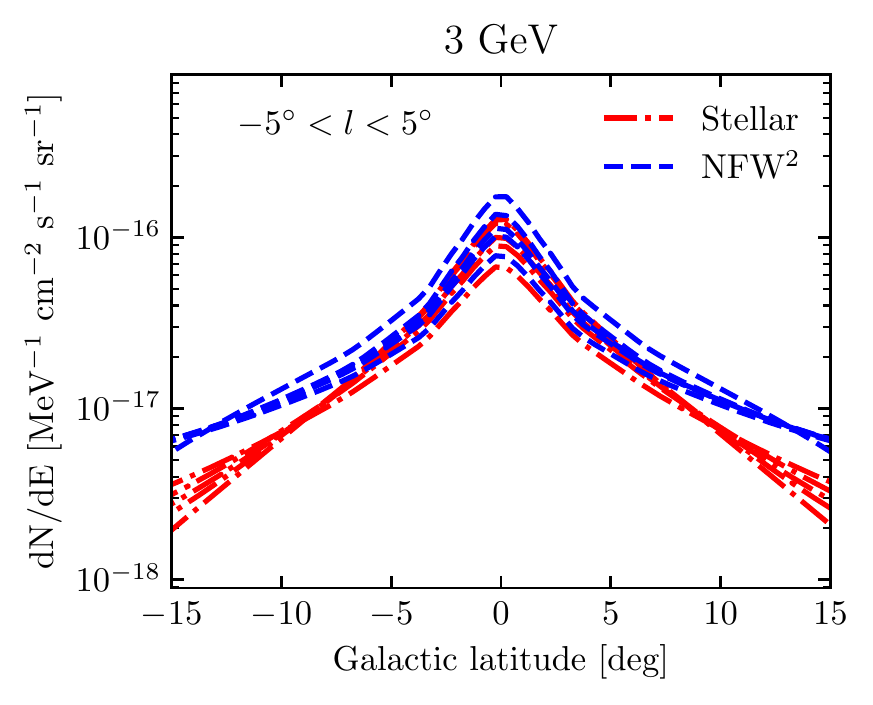}
	\includegraphics[width=0.68\columnwidth]{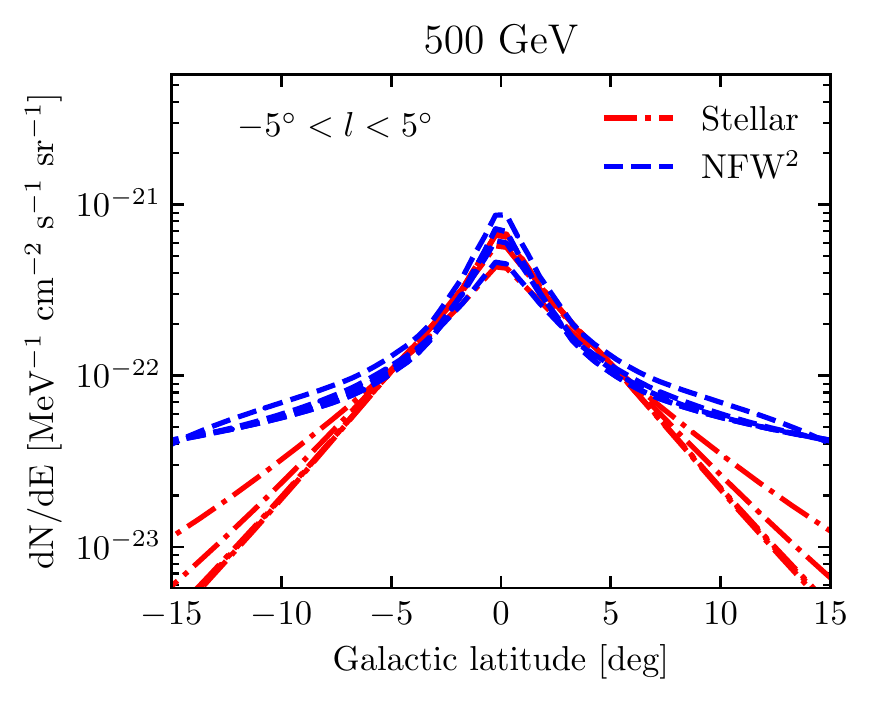}
	\includegraphics[width=0.68\columnwidth]{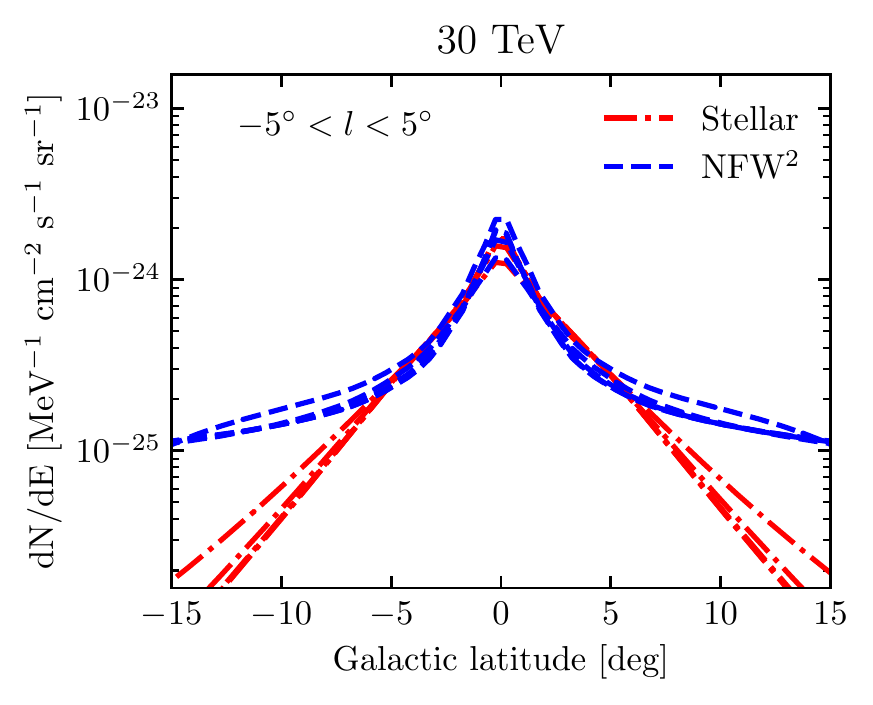}
  \caption{IC longitudinal (top row) and latitudinal (bottom row) profiles for the stellar template (red dot-dashed) and NFW$^2$ template (blue dashed), showing predictions from different propagation models as listed in Table~\ref{tab:dpara}.  The profiles are obtained by summing over 10$^\circ$ wide bands in the latitude and longitude directions, respectively. The columns are for the same energy windows as in Fig.~\ref{fig:skymap_bulge_vs_nfw}. The profiles show consistent changes in energy as seen in the corresponding sky map, and that the differences between the stellar and NFW$^2$ templates are greater than the differences caused by our selection of propagation setup.}
  \label{fig:profile_bulge_vs_nfw}
\end{figure*}

The IC spectra from the Galactic ridge region for different MSP injection spectra (Inj1, Inj2, ..., Inj4) are shown in Fig.~\ref{fig:spectrum_spectral}. We find that most of our predictions are at or below the corresponding H.E.S.S.~data points (assuming $f_{e^\pm} = 10\%$). However, for the hard injection spectrum model Inj1 ($\Gamma = 1.5$ and $E_{\text{cut}}=50$ TeV), the IC spectra overshoots the H.E.S.S.~observations. This means that either the Inj1 model is disfavored, or that $f_{e^\pm}$ is lower than $\sim 6\%$ in the Galactic ridge region. Contributing to this overshooting could be that the $e^\pm$ injection in the NB is overestimated. This is due to the fact that we normalize the $e^\pm$ luminosity by their gamma-ray luminosities. The best-fit NB gamma-ray luminosity obtained by Ref.~\cite{Macias:2016nev} may be somewhat overestimated, because the NB spatial morphology is similar to that of the CMZ structure which could cause spatial degeneracies in the fits. In this sense, a fraction of the gamma-ray photons that are of hadronic origin (emitted by the CMZ) could have been absorbed by the NB stellar template. Also, we note that at around 100 TeV, pair production will attenuate the predicted flux by $\sim 3/4$.

\begin{figure}[t!]
  \includegraphics[width=\columnwidth]{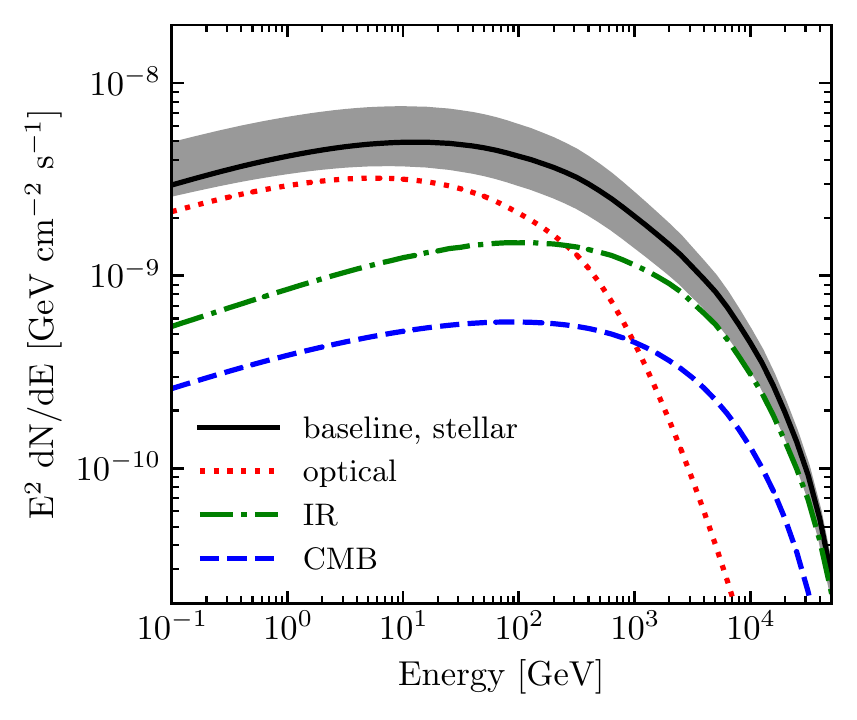}
  \caption{Components of the IC emission for the baseline stellar model from the Galactic ridge, arising from different up-scattered photon fields: optical (red dotted), IR (green dot-dashed), CMB (blue dashed), and the total (black solid). The optical component dominates the GeV energy range, but decreases above 100 GeV, eventually being overtaken by up-scattered IR and CMB photons.}
  \label{fig:spectrum_comp}
\end{figure}

Figure \ref{fig:skymap_bulge_vs_nfw} shows the morphologies of the IC emission for our baseline models in different energy windows: 3 GeV (left), 500 GeV (center), and 30 TeV (right). The top (bottom) row shows the sky maps for the stellar (NFW$^2$) template. The sky maps are normalized by their fluxes at the GC. We find that there are energy-dependent morphological differences between the two IC predictions. These reflect the different $e^\pm$ source distribution models considered. In the GCE energy range ($\sim 3$ GeV, left panels), the IC sky maps are similarly elliptical for both the stellar and NFW$^2$ templates. However, the sky maps become less elliptical at 500 GeV and above. At around $\sim 30$ TeV, the morphologies of the IC component start to show the source distributions displayed in Fig.~\ref{fig:injection_skymap}. As it can be seen, in the highest energy window the sky maps for the stellar template (top-right panel) are boxy while that for the NFW$^2$ template (bottom-right panel) is close to spherical. However, the left-right asymmetry due to the tilt of the bar is not seen in the IC emissions.

These features can be seen more clearly in the corresponding latitudinal and longitudinal profiles presented in Fig.~\ref{fig:profile_bulge_vs_nfw}. Here, we also show the variations in predicted IC morphologies when the propagation setups are varied, as in Table~\ref{tab:dpara}. We note that the morphological differences between the stellar and NFW$^2$ templates are robust to changes in the propagation parameters. It is clear that at tens of TeV, the IC sky maps are sensitive to the source injection distributions.

\begin{figure}[t!]
  \includegraphics[width=\columnwidth]{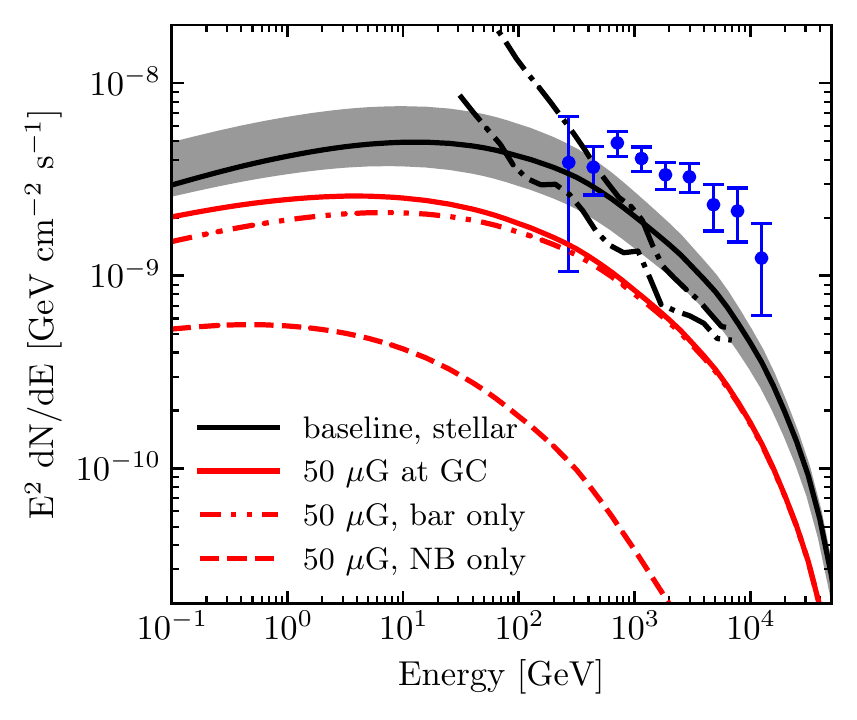}
  \caption{IC emissions from different magnetic field models. The baseline (black solid curve) adopts the original \texttt{GALPROP} magnetic field. The red solid curve introduces a 50 $\mu$G lower limit for the inner 400 pc region around the GC. With the presence of such a lower limit, the IC emission from the Galactic ridge is reduced by $\sim 1/2$. Due to the higher overlap with the modified magnetic field, the IC emission from the NB (red dashed curve) is suppressed more strongly by synchrotron loss compared to the Galactic bar (red dot-dot-dashed curve).}
  \label{fig:spectrum_bfields}
\end{figure}

At IC photon energies of $\sim 1$ GeV, the IC emission is in the nonrelativistic (Thomson) regime and the IC flux is dominated by up-scattered optical photons (Fig.~\ref{fig:spectrum_comp}). The optical photons are mainly emitted by stars whose density peaks along the Galactic plane. As a result, in this energy range the IC emissions are elongated along the Galactic plane and have marked elliptical appearances. This is almost invariant of the spatial morphology of the MSP distributions assumed, explaining the similarity between the left panels of Fig.~\ref{fig:skymap_bulge_vs_nfw}. However, at higher energies, the IC emission starts to enter the relativistic scattering (Klein-Nishina) regime and becomes suppressed, causing the decline of the up-scattered optical photon signal from around 100 GeV (red dotted curve in Fig.~\ref{fig:spectrum_comp}). Since the Klein-Nishina regime is reached at higher energies for lower-energy target photons~\cite{Fermi-LAT:2014sfa}, the IR and CMB photons continue, overtaking the optical photons. Furthermore, as the IR is less concentrated along the Galactic plane than the optical photons (and the CMB is isotropic), the IC emission retains more morphological information of the injected $e^\pm$.

A larger magnetic field means greater energy loss due to synchrotron radiation of the injected $e^\pm$ and a corresponding reduction in the IC emission. We tested a constant magnetic field $B = 50\ \mu$G in the GC region on scales of 400 pc (see Sec.~\ref{sec:bfield}) to explore its impact on our model predictions. Figure~\ref{fig:spectrum_bfields} shows the predicted IC spectrum for the baseline model assuming the default magnetic field (black solid curve with gray shaded band) and the corresponding IC spectrum for the same model but assuming the modified magnetic field (red solid curve). As can be seen, the IC spectrum normalization for the enhanced magnetic field setup is $\sim 1/2$ of the default one. This represents a change in normalization that is larger than our estimated modeling uncertainties from different propagation models (gray shaded). This reduction is mainly due to synchrotron energy loss of $e^\pm$ in the NB. Figure~\ref{fig:spectrum_bfields}  displays the predicted IC spectrum for the NB component (red dashed curve) and Galactic bar (red dot-dot-dashed curve) separately in the enhanced magnetic field case. We observe that for this case the Galactic bar emission dominates. This is different to our predictions obtained using the default magnetic field setup, where the NB and bar  contributions were comparable (see Fig.~\ref{fig:spectrum_diffuse}). This can be understood by noticing that the NB resides in the region where the modified magnetic field underwent the normalization increase. Consequently, much of the $e^\pm$ emitted from this region endure maximum energy loss via synchrotron radiation.

\section{\label{sec:conclu}Discussion and Conclusion}

Recent analyses of the GCE~\cite{Macias:2016nev,Bartels:2017vsx,Macias:2019omb} have revealed its nonspherical nature and have provided further support for an MSP origin. We have revisited the computation of secondary IC emission from the $e^\pm$ injected by such MSPs. Compared to previous studies that assumed a spherically symmetric spatial distribution of MSPs, we adopted 3D models of the stellar distributions in the GC and numerically calculated the IC emissions using the {\tt GALPROP} code. Furthermore, we systematically explored the impact of diffusion parameter uncertainties with additional {\tt GALPROP} runs. We found that the predicted IC fluxes beyond 100 GeV are within the forecasted sensitivity limits of future gamma-ray telescopes for our baseline parameters (Fig.~\ref{fig:spectrum_diffuse}). The very high-energy IC emission from MSPs is nondegenerate with that caused by DM annihilation, from which the $e^\pm$ can only reach a few tens of GeV if DM is responsible for the GCE.

Although the IC spectra from the GC provide insufficient information for identifying the spatial model of the source, we found that the spatial morphology of the IC could serve as a discriminant between the spherically symmetric and 3D stellar distribution injection models (Fig.~\ref{fig:skymap_bulge_vs_nfw}). In the GeV energy range, the IC morphologies are equally elliptical for both the stellar and NFW$^2$ models. However, above $\sim$ TeV energies, they reveal morphological differences that trace the injection distributions. They can therefore be used to discriminate the spherically symmetric and 3D stellar injection models.

Our predicted IC fluxes contribute $\lesssim 10 \%$ of the GCE emission and are at or below than the H.E.S.S.~observations of the Galactic ridge at around a  few TeV. These are consistent with the null detection of secondary emissions in the GeV range \cite{Lacroix:2015wfx} and the dominantly hadronic origins of the H.E.S.S.~measurements ~\cite{Gaggero:2017jts,Abramowski:2016mir}. Thus they constitute important consistency checks of the MSP scenario for the GCE. We compared the IC fluxes with the CTA sensitivity from Ref.~\cite{Silverwood:2014yza} and found that the IC emission could be detected and potentially reveal a signature of GC MSPs with a specialized spectral and morphological search. The HAWC telescope~\cite{DeYoung:2012mj,Abeysekara:2013tka} operates in similar energy bands, and while not having a full view of the Galactic bulge region may have sensitivity to hard $e^\pm$ injection models. To this end, a wide field-of-view TeV gamma-ray observatory in the southern hemisphere is warranted \cite{Mostafa:2017fza}.

The detectability of the IC emission depends on the setup, including MSP $e^\pm$ spectrum and propagation parameters. We adopted a canonical $e^\pm$ power-law slope of 2.0, but softer spectra would make it a challenge to detect the IC component at TeV energies (Fig.~\ref{fig:spectrum_spectral}). Furthermore, an increased magnetic field at the GC would reduce the IC emission via enhanced synchrotron energy losses. In particular, assuming a constant magnetic field of magnitude 50 $\mu$G in the inner 400 pc of the GC, we found a reduction of $\sim 1/2$ of the IC emission obtained with the default magnetic field setup. On the other hand, our model predictions were not very sensitive to changes in the propagation parameters within the 95\% credible contours provided in the 2D marginalized posterior distributions of Ref.~\cite{Johannesson:2016rlh}.

There are various assumptions in our calculations that warrant future detailed studies. For example, we used the default ISRF from the {\tt GALPROP} version 54, which is 2D after averaging the angular dependence. This has recently been updated in Refs.~\cite{Porter:2017vaa,Johannesson:2018bit}, where a 3D ISRF was adopted in the context of CR propagation and high-energy gamma-ray emissions in the Galaxy. Their results show nontrivial impacts from employing the 3D ISRF on the propagation parameters of CRs and the gamma-ray intensity maps. Our results may be affected in many ways, including the fluxes and sky maps. Studies with the {\tt PICARD} code show that new ISRFs increase gamma-ray intensities from the Galactic center, in particular at energies of $\sim 200 $ GeV \cite{Niederwanger:2018zsv}. Note however that our Galactic bar parametrization remains consistent with the ISRF of {\tt GALPROP}. Even though more recent analyses suggest a larger tilt angle, we found that the left-right bulge asymmetry caused by the tilt is washed out in the IC sky maps, and is certainly smaller than the uncertainty caused by propagation parameters.

For the propagation part, we have adopted the results of a wide scan of propagation parameters \cite{Johannesson:2016rlh}. However, caution must be exercised since the propagation properties around the Galactic center may be unique. We also have not covered all possibilities, e.g., we did not consider the possibility of cosmic rays advected out of the region by large-scale outflows. Such outflows may be related to the Fermi bubbles \cite{Crocker:2014fla} and depending on the velocity would affect the secondary IC morphology. We have also neglected MSPs in the Galactic disk, which would provide additional $e^\pm$ injection and IC emission. However, population syntheses show that the MSP contribution to the Galactic diffuse gamma-ray emission is at the few-percent level or less \cite{Gonthier:2018ymi} and we do not expect this to substantially affect our results.

We have modeled the MSP population in the Galactic bulge. However, the presence of younger pulsars and the evolution of the pulsar population were not considered. It has been shown that TeV halos from younger pulsars can contribute to TeV emissions~\cite{Hooper:2017rzt}. This may potentially change the spectral property of IC emissions from the NB where active star formation is ongoing.

The magnetic field at the GC is a crucial parameter affecting the IC emission from a putative MSP population in the nuclear bulge. We have shown that an enhanced magnetic field in this region in turn augments the synchrotron energy losses of the MSP $e^\pm$, thus decreasing the IC yields. The estimated magnetic fields at $\sim$ 100 pc around the GC has large uncertainties and vary from 10 $\mu$G~\cite{LaRosa:2005ai} to 1000 $\mu$G~\cite{Morris1989fd}. Here we only tested the original \texttt{GALPROP} model ($\sim 10\ \mu$G at the GC) and a 50 $\mu$G lower limit obtained in Ref.~\cite{Crocker:2010xc}. An even larger magnetic field at the GC means that the synchrotron radiation would be dominant, especially for the NB component that resides within the 230 pc region around the GC. The spectrum and morphology of the IC emission from the Galactic ridge would potentially be changed by a strong magnetic field in this region. However, the effects on the larger-scale bar/bulge component are expected to be minor. On the other hand, we have only considered the 2D random magnetic field component. A recent study~\cite{Orlando:2019vmq} showed that the IC spatial maps can be significantly affected when more realistic 3D magnetic fields with both random and ordered components are included. This will apply also in the context of MSP secondary emission but its investigation is beyond the scope of the current study.

The Galactic center of the Milky Way offers a unique window to study novel astrophysical and dark matter signals. We have shown that the TeV energy range offers a new handle on the morphology of putative MSPs in the Galactic bulge responsible for the GeV excess. Telescopes such as CTA and HAWC South can be helpful for detecting these IC emissions and for constraining the origin of the GCE in the future.

\begin{acknowledgments}

We thank Roland Crocker for careful reading of the manuscript and insightful suggestions. We thank Kev Abazajian and  Manoj Kaplinghat for comments on the manuscript. D.S. and S.H. are supported by the U.S.~Department of Energy under Awar No. de-sc0018327. This work was partially supported by the World Premier International Research Center Initiative (WPI Initiative), MEXT, Japan. O.M. acknowledges support by JSPS KAKENHI Grant Numbers JP17H04836, JP18H04340 and JP18H04578. The authors acknowledge Advanced Research Computing at Virginia Tech for providing computational resources and technical support that have contributed to the results reported within this paper. URL: \url{http://www.arc.vt.edu}

\end{acknowledgments}

\bibliography{main}
\end{document}